\documentclass[10pt,a4paper,twocolumn]{article}
\usepackage[utf8]{inputenc}
\usepackage[T1]{fontenc}
\usepackage[a4paper,margin=2.5cm]{geometry}
\usepackage{amsmath,amssymb,amsfonts}
\usepackage{algorithmic}
\usepackage{graphicx}
\usepackage{textcomp}
\usepackage{xcolor}
\usepackage{csquotes}
\usepackage{authblk}
\usepackage{pgfgantt}
\usepackage{float}
\usepackage{caption}
\usepackage[colorlinks=true,linkcolor=blue,citecolor=blue,urlcolor=blue]{hyperref}
\usepackage[backend=biber, style=ieee, isbn=false, sortcites, maxbibnames=6, minbibnames=1]{biblatex}

\DefineBibliographyStrings{english}{%
  andothers = {et\addabbrvspace al\adddot}
}
\DefineBibliographyStrings{english}{
  url = {\adddot\space[Online]\adddot\space Available:}
}
\DefineBibliographyStrings{english}{
  urlseen = {Accessed:}
}
\DefineBibliographyStrings{english}{
  pages = {pp\adddot},
  page = {p\adddot}
}

\def\BibTeX{{\rm B\kern-.05em{\sc i\kern-.025em b}\kern-.08em
    T\kern-.1667em\lower.7ex\hbox{E}\kern-.125emX}}

\begin{document}
\title{\textbf{Implementing and Evaluating Post-Quantum DNSSEC in CoreDNS}
}

\author[1]{Julio Gento Suela}
\author[1]{Javier Blanco-Romero}
\author[1]{Florina Almenares Mendoza}
\author[1]{Daniel Díaz-Sánchez}

\affil[1]{Telematic Engineering Department, University Carlos III of Madrid, Leganés, Madrid, Spain}

\renewcommand\Authands{ and }
\renewcommand\Affilfont{\itshape\small}

\maketitle

\begin{abstract}
The emergence of quantum computers poses a significant threat to current secure service, application and/or protocol implementations that rely on RSA and ECDSA algorithms, for instance DNSSEC, because public-key cryptography based on number factorization or discrete logarithm is vulnerable to quantum attacks. This paper presents the integration of post-quantum cryptographic (PQC) algorithms into CoreDNS to enable quantum-resistant DNSSEC functionality. We have developed a plugin that extends CoreDNS with support for five PQC signature algorithm families: ML-DSA, FALCON, SPHINCS+, MAYO, and SNOVA. Our implementation maintains compatibility with existing DNS resolution flows while providing on-the-fly signing using quantum-resistant signatures. A benchmark has been performed and performance evaluation results reveal significant trade-offs between security and efficiency. The results indicate that while PQC algorithms introduce operational overhead, several candidates offer viable compromises for transitioning DNSSEC to quantum-resistant cryptography.
\end{abstract}

\date{}

\section{Introduction}

From ancient times, knowledge has been established as one of the most valuable assets. This gave rise to the need for it to be transmitted, but at the same time, it needed to be protected from falling into the wrong hands~\cite{importanciacrip}.
Cryptography emerged in response to this need, transforming information in such a way that only those with knowledge of a key could unveil the content of encrypted messages.

Currently, the most widely used cryptographic algorithms, such as RSA (\textit{Rivest, Shamir, and Adleman}) and ECC (\textit{Elliptic Curve Cryptography}), could become vulnerable to attacks carried out with quantum computers, which calls into question their long-term viability in guaranteeing data security.

This drives us towards the development of new cryptographic schemes and thus to an advancement in quantum computing. In recent years, research in this field has intensified~\cite{PQCexpRIPE}, leading organizations like the National Institute of Standards and Technology (NIST) to warn about the risk that quantum computers pose to current encryption systems.
Although it is not known exactly when this will happen, many experts estimate that it could occur in the next 10 years, which makes the transition to post-quantum cryptography (PQC) urgent, given that the implementation of new cryptographic infrastructures can take decades. NIST began standardizing quantum-resistant algorithms in 2016, with the aim of ensuring they are also secure against classical attacks and compatible with current systems~\cite{nistpqc}.

As a result of this effort, on August 13, 2024, NIST published the ML-KEM (\textit{Module-Lattice–based Key Encapsulation Mechanism})~\cite{fips203}, ML-DSA (\textit{Module-Lattice–based Digital Signature Algorithm})~\cite{fips204}, and SLH-DSA (\textit{Stateless Hash-based Digital Signature Algorithm})~\cite{fips205} standards based on the CRYSTALS-KYBER, CRYSTALS-ML-DSA, and SPHINCS+ algorithms, respectively. Additionally, it selected and is preparing the FN-DSA standard, based on the FALCON algorithm for digital signature. Subsequently, on March 2025, the selection of the HQC algorithm for key exchange was announced, as an alternative to ML-KEM, for standardization~\cite{nistpqc}. Furthermore, in 2024, NIST initiated a new phase of the standardization process focused on additional post-quantum digital signature algorithms. This phase, referred to as the second round of proposals for digital signature algorithms, seeks to explore complementary options, evaluating new candidates such as MAYO, SNOVA, CROSS, etc. These algorithms could offer improvements in efficiency or versatility compared to the already standardized schemes and are being considered for possible future standardization~\cite{nist2024signatures2ndround}. These initiatives are part of a continuous effort to ensure cryptographic security in the post-quantum era.

Thus, these new cryptographic algorithms represent a fundamental step towards cryptographic security in a world with quantum computing, but they also pose significant operational challenges, especially in communication protocols. In this context, we focus specifically on DNSSEC (\textit{Domain Name System Security Extensions})~\cite{rfc4033,rfc4034,rfc4035}, which provides authentication and data integrity for DNS responses through a hierarchical chain of digital signatures, protecting against attacks such as cache poisoning and DNS spoofing. However, current DNSSEC deployments rely primarily on RSA and ECDSA algorithms~\cite{rfc5702}. Some DNS servers have experimental support of PQ algorithms, for instance BIND9, developed by the Internet Systems Consortium (ISC) and one of the most widely used DNS software implementations, but in most of them are not supported yet. Thus, the transition to post-quantum cryptography in DNSSEC presents unique challenges, particularly regarding the size of post-quantum signatures and keys, which in some cases can be between 10 and 50 times larger than those generated with traditional algorithms. This increase in size not only impacts the efficient transmission of data in UDP-based (\textit{User Datagram Protocol}) networks, such as DNS (\textit{Domain Name System}), but also increases memory requirements for servers and DNS resolvers, often forcing the use of TCP transport~\cite{rfc7766} which introduces additional latency~\cite{verisignpqdnssec}.
 
Before large-scale deployment, testing tools are needed to evaluate the performance and compatibility impact of integrating these new algorithms into real systems. Controlled testing environments help identify potential issues and ensure that migration to PQ algorithms does not compromise system security or operability~\cite{PQCexpRIPE}.

While initial research has explored PQC integration in specific DNS implementations, broader evaluation across different DNS server architectures remains needed. In this paper, we describe the integration of standard and candidates, in the second NIST round for signatures, of post-quantum algorithm families (FALCON, ML-DSA, SPHINCS+, MAYO, and SNOVA) into CoreDNS, a widely-used DNS server in Kubernetes environments developed in Go. This implementation extends existing work by providing empirical performance data across a broader range of PQC algorithms, enabling comparative analysis for algorithm selection and deployment planning in containerized DNS environments.

The rest of this paper is organized as follows. Section~\ref{sec:soa} explains the related and previous work. In section~\ref{sec:solution}, we present system requirements, proposed architecture to carry them out and the implementation details. Section~\ref{sec:results} shows the tests that validate the functioning of the proposed implementation and their respective analyses. Finally, section~\ref{sec:conclusions} presents the final conclusions and future lines to extend this work.

\section{Related Work}
\label{sec:soa}

The evolution of cryptographic algorithms in DNSSEC has followed a progression from weaker to stronger classical algorithms, as documented in RFC 8624~\cite{wouters2019algorithm}, which established current algorithm implementation requirements. The standard recommends ECDSAP256SHA256 as the primary algorithm, with ED25519 as the emerging default, while marking RSA-based algorithms for gradual deprecation. However, all recommended algorithms remain vulnerable to quantum attacks, creating the need for post-quantum alternatives.

Previous research by van Rijswijk-Deij et al.~\cite{van2015making,van2016performance} demonstrated that elliptic curve cryptography could address DNSSEC's packet size and performance issues compared to RSA, showing that ECC signature validation, while slower than RSA, remained manageable for DNS resolvers. This work established the foundation for algorithm transitions in DNSSEC and highlighted the performance trade-offs inherent in cryptographic upgrades.

The first systematic analysis of post-quantum cryptography integration in DNSSEC was conducted by Müller et al.~\cite{muller2020retrofitting}, who identified three potentially suitable candidates (Falcon-512, Rainbow-Ia, and RedGeMSS128) through theoretical analysis. Their work revealed that protocol modifications would be necessary due to increased signature sizes and established the 1,232-byte UDP limit as a critical constraint. This theoretical foundation was extended by Müller's thesis~\cite{muller2021making}, which identified signature size as the most significant deployment barrier and proposed out-of-band key distribution as a potential solution.

Parallel work by Shrishak and Shulman~\cite{shrishak2021negotiating} addressed the deployment challenge through cipher-suite negotiation mechanisms, recognizing that DNSSEC's lack of algorithm negotiation creates adoption barriers when performance trade-offs are significant. Their proposed end-to-end and hop-by-hop negotiation approaches would enable dynamic algorithm selection based on resolver capabilities.

Practical implementation began with Beernink's thesis~\cite{beernink2022taking}, which validated theoretical predictions using real DNS traffic traces from University of Twente and .nl ccTLD logs. The work confirmed that only Falcon-512 could be adopted without protocol modifications and implemented an out-of-band key exchange solution in Unbound, demonstrating approximately 30\% overhead but enabling better-performing algorithms.

Technical solutions for packet size limitations emerged with Rawat and Jhanwar's~\cite{rawat2023post} QNAME-based fragmentation (QBF) approach, which addressed the UDP size constraints through application-layer fragmentation. Testing with Falcon-512, ML-DSA-44, and SPHINCS+ showed that QBF could resolve queries in 43±1ms compared to 83±1ms for standard DNS with TCP fallback, using only standard DNS resource records.

Recent empirical evaluation has expanded significantly. Schutijser et al.~\cite{schutijser2024testbed} developed the PATAD testbed using PowerDNS to create a testing infrastructure for PQC algorithms across complete DNS hierarchies. Their work integrates Falcon-512 and evaluates MAYO-1 and SQISign-1 in container-based environments simulating real deployment scenarios.

Field testing reached production scale with Goertzen et al.'s~\cite{PQCexpRIPE} PowerDNS study, which deployed PQC-signed zones across the RIPE ATLAS network of approximately 10,000 global nodes. The study confirmed that delivery rates decrease with signature size, with Falcon experiencing fewer issues than SPHINCS+ and XMSS. Notably, 8.5\% of resolvers incorrectly indicated successful validation for Falcon signatures despite lacking PQC support, revealing implementation challenges.

Current operational evaluation by Schutijser et al.~\cite{schutijser2025evaluating} focuses on TLD operator requirements using real zone files from .nl, .se, and .nu domains. Their results demonstrate that MAYO-2 outperforms RSA-1280 in signing performance while Falcon-512 performs comparably, with hardware acceleration showing significant improvements on x86-64-v3 compared to v2 architectures.

Another relevant effort is OQS-BIND~\cite{oqsbind2025}, which enables experimental usage of PQC algorithms in a BIND 9 fork, supporting Falcon-512, ML-DSA-44, and SPHINCS+-SHA-256-128s through liboqs integration. As mentioned before, BIND, developed by the Internet Systems Consortium, is one of the most widely used DNS software implementations. 

Despite this progress, production-ready PQC implementations in DNS servers remain limited. Current research focuses on three main areas: algorithm performance optimization, protocol adaptation for larger signatures, and deployment strategy development. The present work addresses this gap by implementing five PQC algorithm families in CoreDNS, providing further empirical performance data for algorithm selection and deployment planning.

\section{System Description}
\label{sec:solution}

This section describes the architecture and implementation of our post-quantum DNSSEC solution for CoreDNS. We present the CoreDNS plugin-based architecture, our selection of post-quantum cryptographic algorithms, and the \texttt{dnssec\_pqc} plugin implementation that enables quantum-resistant DNS signing.

\subsection{CoreDNS}

CoreDNS holds an increasingly important position within the modern DNS ecosystem, primarily due to its role as the default DNS server for Kubernetes clusters, and its design and capabilities, which are particularly well-suited for dynamic, cloud-native, and microservices-based environments.

CoreDNS is a DNS server developed in the Go programming language, characterized by a plugin-based architecture sequence of plugins to be executed are defined~\cite{coredns2025, coredns2025git}. This structure allows for high flexibility, as each plugin can independently fulfill various DNS functions. The server is configured through a file called the Corefile, which defines server blocks, DNS zones, ports, and the sequence of plugins to be executed.

When CoreDNS receives a DNS query, it identifies the appropriate server block based on the most specific zone match. The query then passes through the associated chain of plugins, where each plugin can either respond directly, delegate to the next plugin, apply a fallthrough mechanism, or provide auxiliary information. Processing continues until a plugin generates a response or all plugins have been exhausted, resulting in an error.

CoreDNS includes approximately 30 default plugins covering basic functionality such as logging, caching, and metrics collection, while also supporting external plugin integration. The server operates with multiple protocols including traditional DNS, DNS over TLS, DNS over HTTPS, and gRPC. Notable among its plugins is \texttt{dnssec}, which enables on-the-fly signing of DNS responses using cryptographic algorithms such as RSA, ECDSA, or ED25519.

\subsection{Signed DNS Query Flow}

The flow of a DNS query signed with conventional algorithms in CoreDNS begins with the configuration of the Corefile, in which a specific zone (for example, {\tt mysig.com}) is defined and the \texttt{dnssec} plugin is included. This plugin dynamically signs DNS responses with RRSIG records using locally stored keys.

During startup, CoreDNS parses the Corefile, loads the plugins (such as \texttt{dnssec}, \texttt{forward}, and \texttt{log}), and waits for queries on the configured port. When a client makes a query (for example, using \texttt{dig}), it is processed by the plugin chain. If the response is locally available, CoreDNS generates the DNSSEC signature in real time for the record sets (RRsets), adding an RRSIG record with cryptographic metadata such as algorithm, validity period, key used, and the signature itself.

\subsection{PQ Cryptographic Requirements}

Once the internal functioning of CoreDNS was understood, its digital signature system was modified to incorporate post-quantum cryptographic algorithms. The selection of PQC algorithms was based on criteria such as efficiency, key/signature size, and security level. 

PQ algorithms that are standard or are being analyzed by the NIST with suitable signature and/or key sizes were selected and integrated. Such algorithms with their public key, private key, and signature sizes are the following ones:

\textbf{ML-DSA} provides higher security at the cost of larger keys and signatures, making it useful for comparing performance impacts.

\begin{table}[htbp]
\caption{ML-DSA Algorithm Schemes \cite{liboqs}}
\centering
\resizebox{\columnwidth}{!}{%
\begin{tabular}{|l|c|c|c|}
\hline
\textbf{Algorithm} & \textbf{Public} & \textbf{Private} & \textbf{Signature} \\
\textbf{Schemes} & \textbf{Key (B)} & \textbf{Key (B)} & \textbf{(B)} \\
\hline
ML-DSA-44  & 1312 & 2560 & 2420 \\
\hline
ML-DSA-65 & 1952 & 4032 & 3309 \\
\hline
ML-DSA-87 & 2592 & 4896 & 4627 \\
\hline
\end{tabular}%
}
\label{tab:mldsa}
\end{table}

\textbf{FALCON} offers reasonable signature and key sizes, with "padded" variants that enhance side-channel resistance.

\begin{table}[htbp]
\caption{Falcon Algorithm Schemes \cite{liboqs}}
\centering
\resizebox{\columnwidth}{!}{%
\begin{tabular}{|l|c|c|c|}
\hline
\textbf{Algorithm} & \textbf{Public} & \textbf{Private} & \textbf{Signature} \\
\textbf{Schemes} & \textbf{Key (B)} & \textbf{Key (B)} & \textbf{(B)} \\
\hline
Falcon-512 & 897 & 1281 & 752 \\
\hline
Falcon-1024 & 1793 & 2305 & 1462 \\
\hline
Falcon-padded-512 & 897 & 1281 & 666 \\
\hline
Falcon-padded-1024 & 1793 & 2305 & 1280 \\
\hline
\end{tabular}%
}
\label{tab:falcon}
\end{table}

\textbf{SPHINCS+}, although the signatures are large, the keys are small. Balanced variants with different hash functions were chosen.

\begin{table}[htbp]
\caption{SPHINCS+ Algorithm Schemes \cite{liboqs}}
\centering
\resizebox{\columnwidth}{!}{%
\begin{tabular}{|l|c|c|c|}
\hline
\textbf{Algorithm} & \textbf{Public} & \textbf{Private} & \textbf{Signature} \\
\textbf{Schemes} & \textbf{Key (B)} & \textbf{Key (B)} & \textbf{(B)} \\
\hline
SPHINCS+-SHA2-128s & 32 & 64 & 7856 \\
\hline
SPHINCS+-SHAKE-128s & 32 & 64 & 7856 \\
\hline
\end{tabular}%
}
\label{tab:sphincs}
\end{table}

\textbf{MAYO} offers lightweight signatures with somewhat large public keys.

\begin{table}[htbp]
\caption{MAYO Algorithm Schemes \cite{liboqs}}
\centering
\resizebox{\columnwidth}{!}{%
\begin{tabular}{|l|c|c|c|}
\hline
\textbf{Algorithm} & \textbf{Public} & \textbf{Private} & \textbf{Signature} \\
\textbf{Schemes} & \textbf{Key (B)} & \textbf{Key (B)} & \textbf{(B)} \\
\hline
MAYO-1 & 1420 & 24 & 454 \\
\hline
MAYO-3 & 2986 & 32 & 681 \\
\hline
\end{tabular}%
}
\label{tab:mayo}
\end{table}

\textbf{SNOVA} is one selected scheme that balances security with key and signature sizes.

\begin{table}[htbp]
\caption{SNOVA Algorithm Schemes \cite{liboqs}}
\centering
\resizebox{\columnwidth}{!}{%
\begin{tabular}{|l|c|c|c|}
\hline
\textbf{Algorithm} & \textbf{Public} & \textbf{Private} & \textbf{Signature} \\
\textbf{Schemes} & \textbf{Key (B)} & \textbf{Key (B)} & \textbf{(B)} \\
\hline
SNOVA\_24\_5\_4 & 1016 & 48 & 248 \\
\hline
SNOVA\_24\_5\_4\_SHAKE & 1016 & 48 & 248 \\
\hline
\end{tabular}%
}
\label{tab:snova}
\end{table}

\subsection{\texttt{dnssec\_pqc} Plugin Architecture}

To integrate post-quantum cryptographic algorithms into CoreDNS, we developed the \texttt{dnssec\_pqc} plugin~\cite{gentosuela2025plugin} as an extension of the existing \texttt{dnssec} plugin, maintaining compatibility with CoreDNS's modular architecture while minimizing interference with existing functionality.

The implementation integrates with liboqs 0.14.0-rc1~\cite{liboqs} through Go bindings provided by \verb+liboqs-go+~\cite{liboqsgo}, enabling access to post-quantum cryptographic primitives. This required modifications to the underlying \verb+miekg/dns+ library~\cite{miekgdns2025} in a fork~\cite{gentosuela2025dnspqc} to support new algorithm identifiers and handle the PQC algorithms. The plugin extends CoreDNS's key management system to accommodate PQC keys while maintaining the same configuration interface as the original \texttt{dnssec} plugin.

The core functionality extends the RRSIG record generation process to support PQC algorithms. During DNS response processing, the plugin computes signatures over DNS resource record sets, where a hash is computed over the record set (RRset) using the private key, temporal validity intervals are defined through inception and expiration fields, the signature is linked to its public key via an identifier, and the signing algorithm is specified.

\begin{figure}[H]
    \centerline{\includegraphics[width=0.5\textwidth]{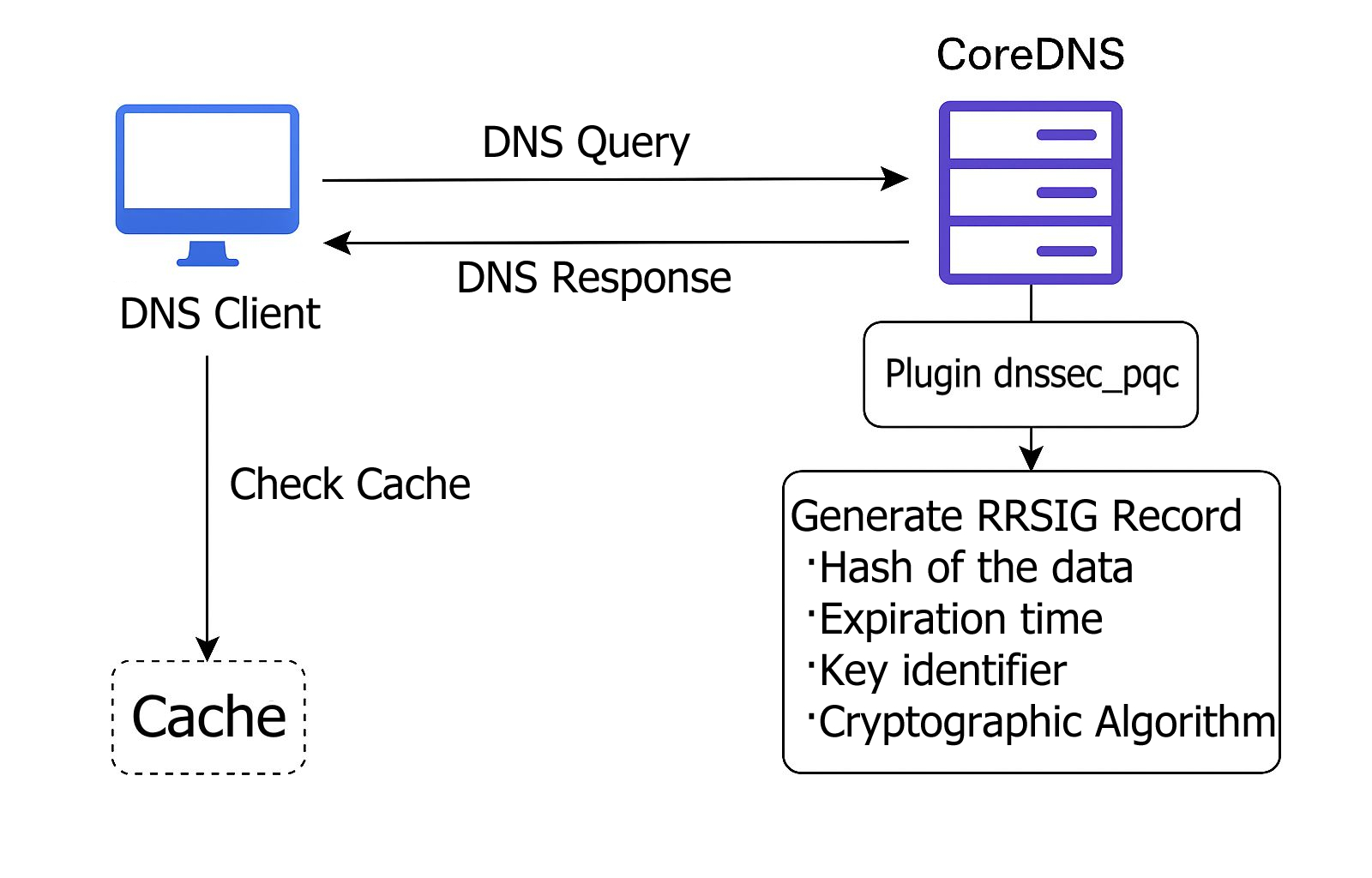}}
    \caption{DNS query flow with post-quantum cryptographic signing in CoreDNS}
    \label{fig:architecture}
\end{figure}

The architecture proposed in Figure \ref{fig:architecture} illustrates the DNS query flow in CoreDNS with the \texttt{dnssec\_pqc} plugin. This system, composed of a DNS client and a CoreDNS-based server, emulates a realistic environment.

\section{Evaluation Results}
\label{sec:results}

To evaluate the functionality and performance of the CoreDNS server with the implemented digital signature algorithms (both classical and post-quantum), functional and performance tests were conducted. The functional tests, performed manually, verified that DNS responses correctly included the expected records (A and RRSIG) signed by each algorithm. The performance tests, automated through the \verb+./scripts/dns_test.sh+ script in the dnssec\_pqc\_plugin repository, measured the impact of each algorithm on server performance, recording metrics such as DNS query resolution latency, CPU cycles, memory usage, DNS response size, and signing time.

To collect the performance metrics, we used several standard Unix tools with specific measurement approaches. The \verb+dig+ tool issues DNS queries and validates the correct signature of records, with query resolution time measured using high-precision timestamps captured before and after command execution, providing microsecond-level accuracy. The \verb+perf+ tool monitors hardware performance counters to track CPU cycles consumed during cryptographic operations by attaching to the CoreDNS server process; a brief sleep period (0.035 seconds) is introduced after \verb+perf+ attachment to allow proper synchronization before DNS query execution begins. Memory usage is captured by monitoring the \verb+/proc/PID/status+ file, specifically extracting the VmHWM value which represents the peak resident set size (RSS) - the maximum amount of physical memory used by the CoreDNS process during operation. To ensure the reliability of the results, each test was repeated 100 times with complete CoreDNS process restart between iterations.

The testing environment consisted of Ubuntu 24.04.2 LTS (Noble Numbat) running on a Dell G15-5510 laptop equipped with an Intel Core i7-10870H CPU @ 2.20GHz processor and 16 GB of RAM and SSD storage.

The evaluation encompassed 18 algorithms: 13 post-quantum variants (Falcon-512, ML-DSA-44, SPHINCS+-SHA2-128s-simple, MAYO-1, Falcon-1024, ML-DSA-65, SPHINCS+-SHAKE-128s-simple, MAYO-3, Falcon-padded-512, ML-DSA-87, Falcon-padded-1024, SNOVA\_24\_5\_4, SNOVA\_24\_5\_4\_SHAKE) and 5 traditional algorithms (RSA-2048, RSA-4096, ECDSA-P256, ECDSA-P384, Ed25519).

The tests confirmed that the integration of post-quantum digital signature algorithms into CoreDNS was successfully carried out. The developed architecture is modular and flexible, facilitating the incorporation of new algorithms in the future. DNS responses signed with post-quantum algorithms are generated and delivered correctly, ensuring compatibility with the name resolution flow.

\subsection{Signing Time}

\begin{figure}[H]
\centerline{\includegraphics[width=1\columnwidth]{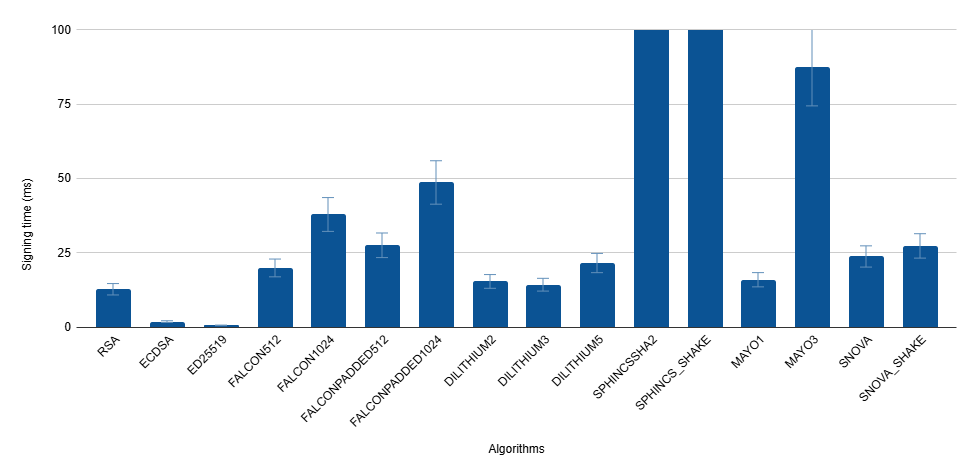}}
\caption{Digital signature generation time for traditional and post-quantum cryptographic algorithms. The error bars show the standard deviation over 10 iterations}
\label{fig:signing_time}
\end{figure}

Figure~\ref{fig:signing_time} presents the digital signature generation times across all evaluated algorithms. Among the traditional algorithms, Ed25519 demonstrates the lowest signing times, closely followed by ECDSA-P256. RSA algorithms exhibit higher signing times with considerable variability, particularly RSA-4096. In the post-quantum domain, results show greater diversity. Schemes like SNOVA, MAYO-1, and ML-DSA-44 offer competitive signing times, averaging below 25 ms. In contrast, Falcon-padded-1024 and SPHINCS+ variants are at the opposite end, with signing times exceeding one second, reaching up to 2.6 seconds for SPHINCS+-SHAKE-128s-simple.

\subsection{Resolution Latency}

\begin{figure}[H]
\centerline{\includegraphics[width=1\columnwidth]{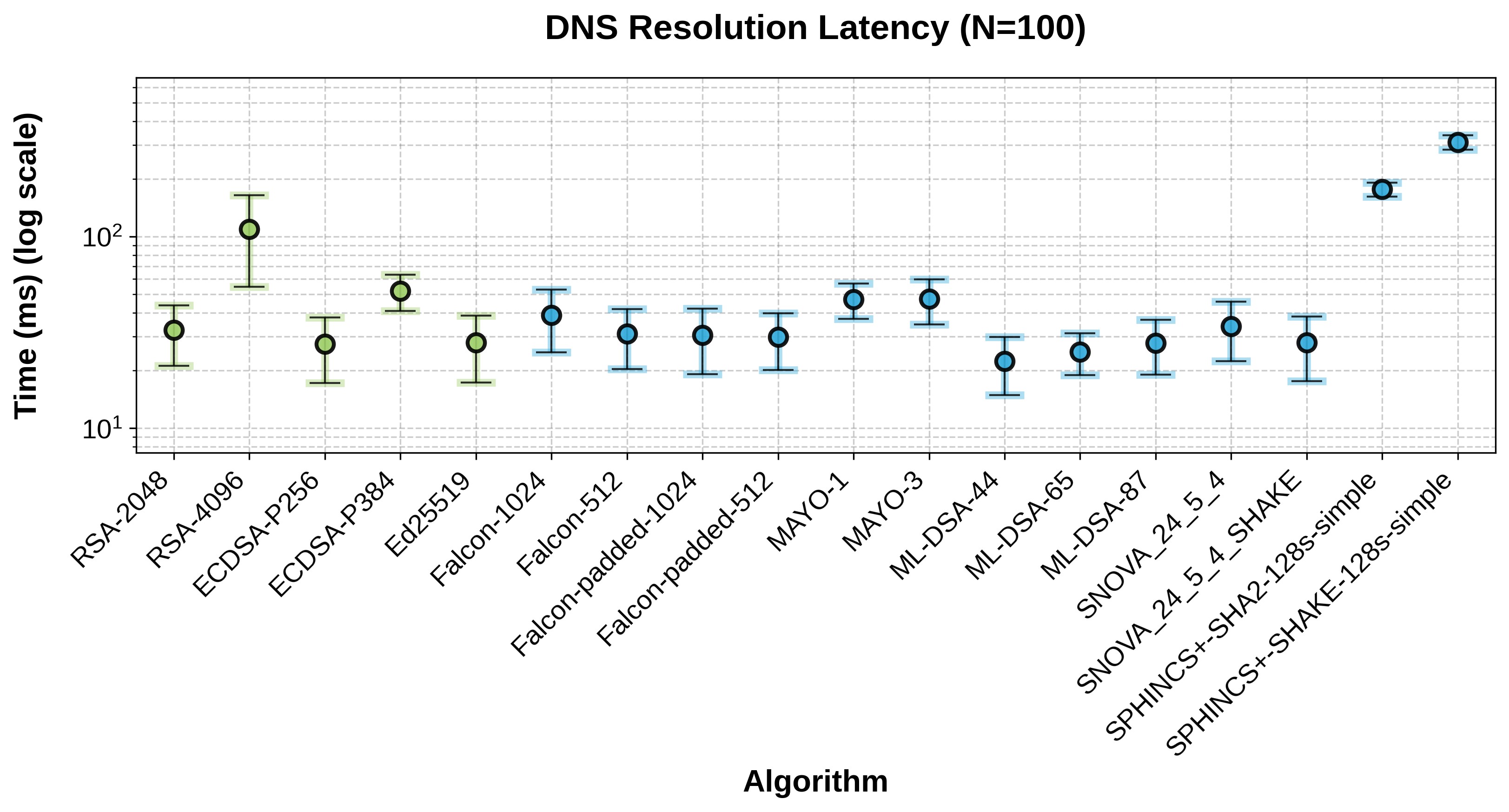}}
\caption{DNS query resolution latency measured from client perspective}
\label{fig:resolution_latency}
\end{figure}

DNS query resolution latency, as shown in Figure~\ref{fig:resolution_latency}, follows a similar pattern to signing time but is also influenced by the need to use TCP due to response size. Traditional algorithms maintain latencies around $10^{1.5}$ ms, with the exception of RSA-4096 which reaches $10^2$ ms. All post-quantum algorithms require TCP, resulting in latencies around $10^{1.5}$ ms for most variants. However, SPHINCS+ algorithms exhibit excessive latencies above $10^2$ ms. Among post-quantum algorithms, MAYO variants show slightly higher latency compared to ML-DSA, Falcon, and SNOVA variants.

\subsection{CPU Usage}

\begin{figure}[H]
\centerline{\includegraphics[width=1\columnwidth]{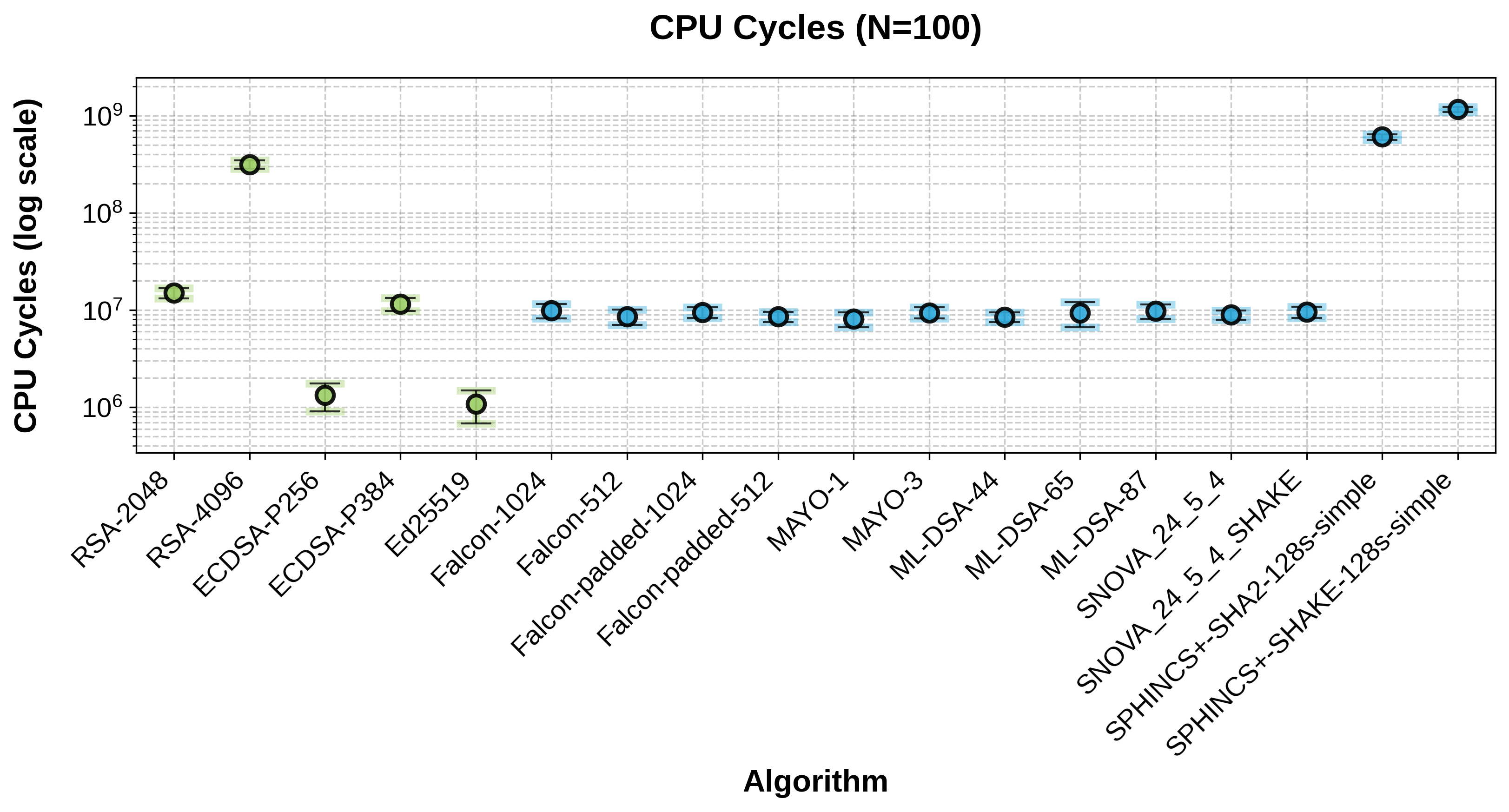}}
\caption{CPU cycles consumed during cryptographic operations measured via hardware performance counters}
\label{fig:cpu_usage}
\end{figure}

CPU usage patterns, illustrated in Figure~\ref{fig:cpu_usage}, reveal distinct performance tiers. ECDSA-P256 and Ed25519 demonstrate the lowest CPU consumption at approximately $10^6$ cycles. RSA-2048 and ECDSA-P384 consume around $10^7$ cycles, similar to most post-quantum algorithms. RSA-4096 falls between $10^8$ and $10^9$ cycles. Most post-quantum algorithms cluster around $10^7$ cycles, with SPHINCS+ variants being notable outliers, consuming nearly $10^9$ cycles, making them the most CPU-intensive algorithms tested.

\subsection{Memory Usage}

\begin{figure}[H]
\centerline{\includegraphics[width=1\columnwidth]{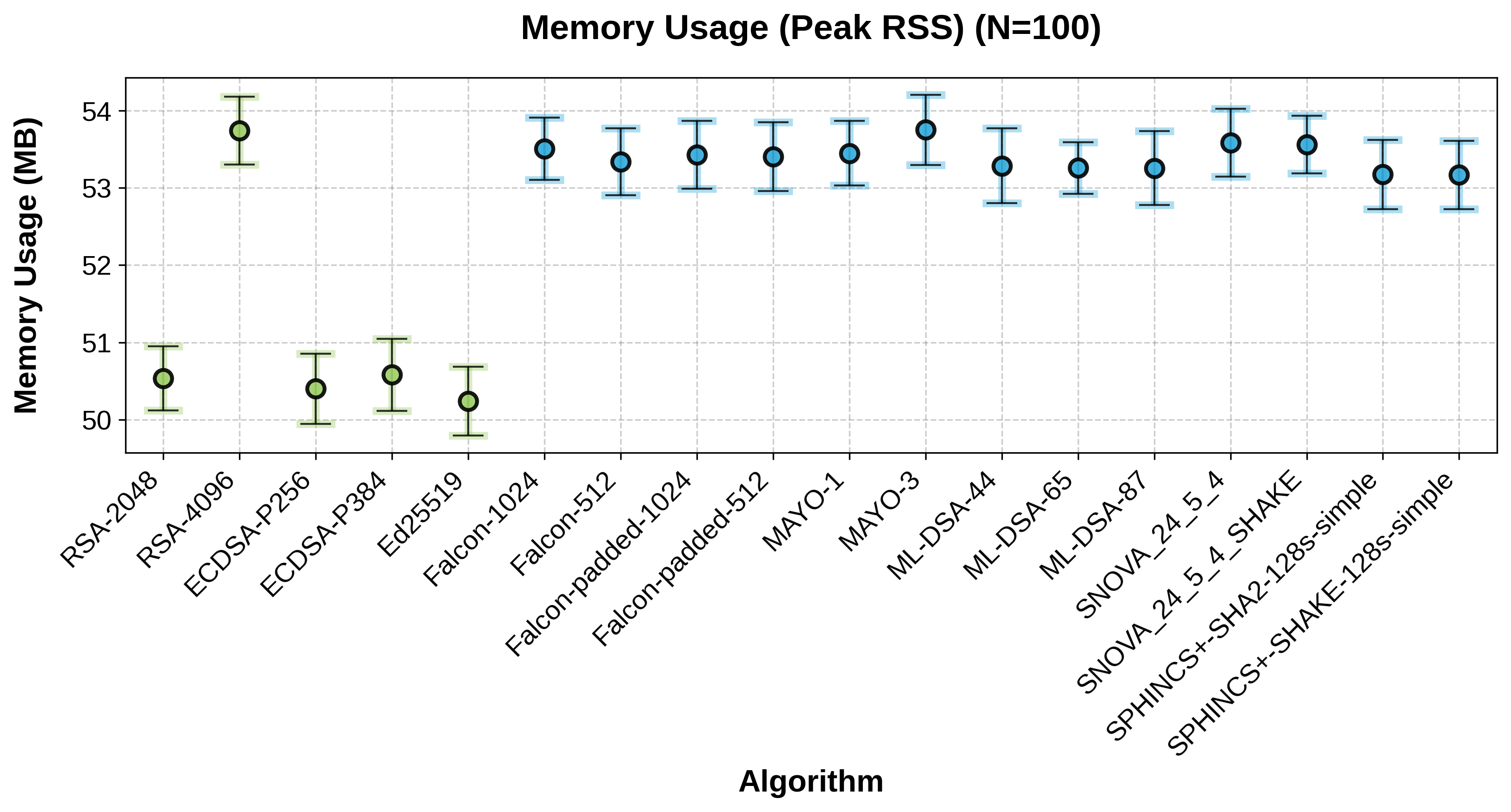}}
\caption{Peak resident set size (RSS) memory consumption of CoreDNS process}
\label{fig:memory_usage}
\end{figure}

Memory consumption, represented as peak RSS in Figure~\ref{fig:memory_usage}, shows relatively stable patterns across algorithm families. Traditional algorithms (excluding RSA-4096) consume between 50-51 MB. All post-quantum algorithms consume between 53-54 MB, together with RSA-4096. The memory overhead for post-quantum algorithms is modest, with an increase of approximately 3-4 MB compared to traditional algorithms. Memory usage shows little variation within algorithm families, suggesting that CPU cycles rather than memory will be the primary limiting factor in high-performance environments.

\subsection{DNS Response Size}

\begin{figure}[H]
\centerline{\includegraphics[width=1\columnwidth]{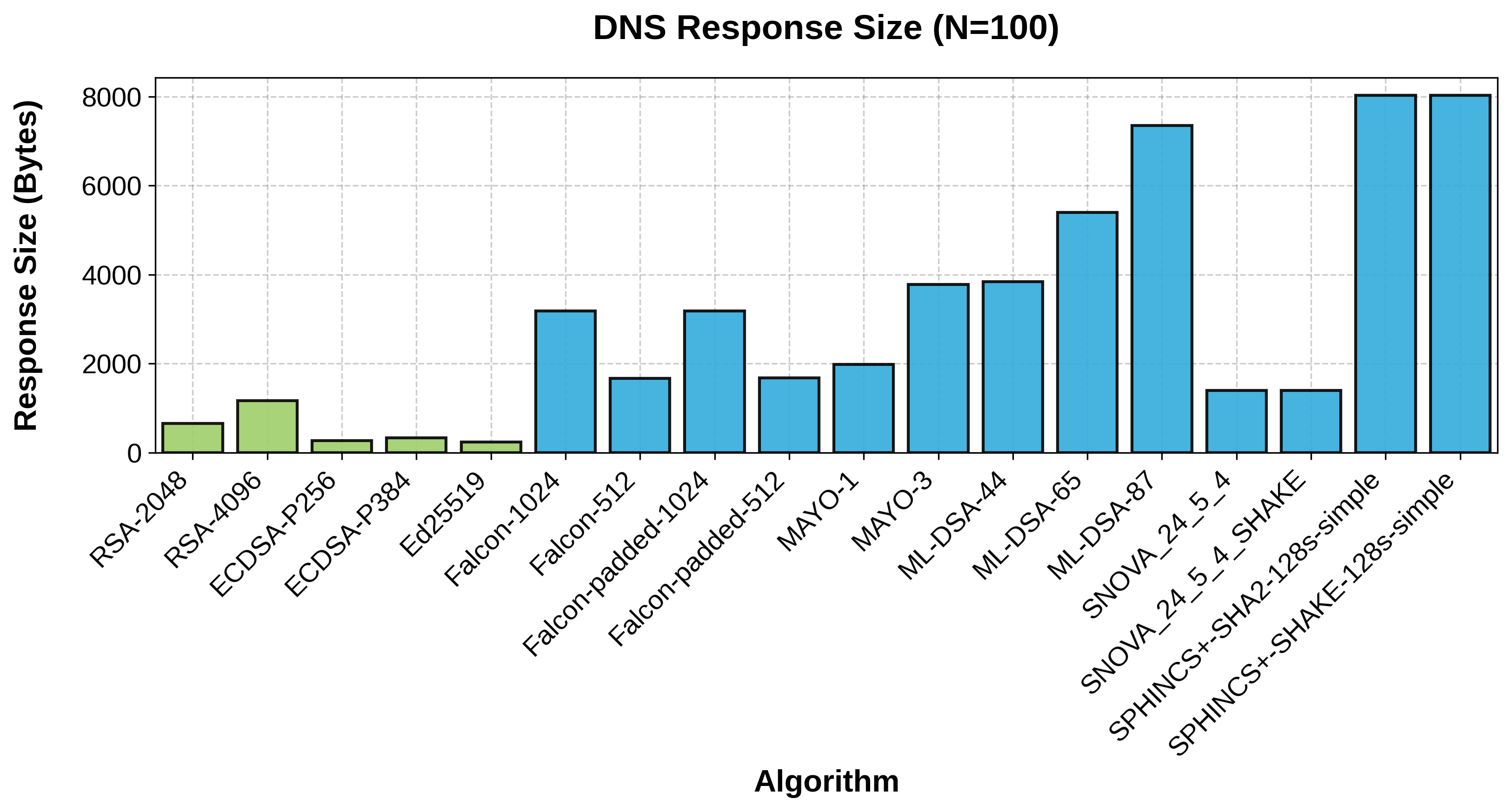}}
\caption{DNS response packet size including DNSSEC signatures for different cryptographic algorithms}
\label{fig:response_size}
\end{figure}

DNS response size, shown in Figure~\ref{fig:response_size}, represents one of the most significant differences between algorithm families. Traditional algorithms generate compact responses that fit within UDP transport limits, resulting in higher efficiency and reduced network overhead. All post-quantum schemes exceed typical UDP size limits, forcing the use of TCP. Within the post-quantum group, SNOVA produces the smallest responses, followed by Falcon-512 and MAYO-1, making them attractive for bandwidth-constrained environments. ML-DSA-87 and Falcon-1024 produce the largest responses, potentially causing fragmentation and interoperability issues in certain networks.

\section{Discussion}
\label{sec:discussion}

The experimental results reveal distinct performance patterns between traditional and post-quantum cryptographic algorithms in DNSSEC implementations. Traditional algorithms maintain superior efficiency across all metrics, with Ed25519 and ECDSA-P256 achieving signing times under 2 ms, CPU consumption around $10^6$ cycles, and compact responses suitable for UDP transport.

Among post-quantum alternatives, Falcon, ML-DSA, MAYO, and SNOVA variants demonstrate practical viability for DNSSEC deployment, with CPU consumption around $10^7$ cycles and signing times between 15-50 ms. Most post-quantum algorithms perform comparably to RSA-2048 and ECDSA-P384, while achieving modest efficiency gains over RSA-4096. SPHINCS+ algorithms present significant operational challenges, with signing times exceeding one second and CPU consumption approaching $10^9$ cycles, limiting their applicability in high-throughput DNS environments.

A fundamental challenge affecting all post-quantum algorithms is the requirement for TCP transport due to response size limitations. This constraint introduces additional latency and complexity compared to traditional UDP-based DNS operations, representing a protocol-level limitation that affects deployment strategies and may necessitate complementary solutions such as response fragmentation techniques or alternative key distribution mechanisms.

\section{Conclusions and Future Work}
\label{sec:conclusions}

This work successfully demonstrates the integration of post-quantum cryptographic algorithms into CoreDNS, achieving functional DNS responses with quantum-resistant signatures. The implementation extends the existing \texttt{dnssec} plugin architecture through a separate \texttt{dnssec\_pqc} plugin, maintaining compatibility with CoreDNS's modular design while minimizing interference with existing functionality.

The performance evaluation of 18 algorithms reveals that while classical algorithms remain the most efficient option, several post-quantum candidates offer practical compromises for DNS deployment. Among PQC options, ML-DSA, SNOVA, MAYO, and Falcon variants demonstrate the most balanced performance characteristics, with signing times between 15-50 ms, CPU usage around $10^7$ cycles, and manageable response sizes. These findings complement results by Schutijser et al.~\cite{schutijser2025evaluating}.

Memory overhead remains modest across all post-quantum algorithms, with only a 3-4 MB increase compared to traditional algorithms. The measured signing times of 15-50ms for viable PQC candidates align with the approximately 30\% overhead observed by Beernink~\cite{beernink2022taking} in Unbound, indicating that performance impact remains within acceptable bounds for DNS operations. This confirms theoretical predictions by Müller et al.~\cite{muller2020retrofitting} and aligns with practical observations from large-scale deployments~\cite{PQCexpRIPE}.

The modular architecture provides flexibility for future algorithm additions, positioning CoreDNS to adapt as post-quantum cryptography standards evolve and supporting cryptographic agility principles. The performance trade-offs between security and efficiency indicate that organizations can begin planning gradual transitions to quantum-resistant mechanisms based on specific deployment requirements. The need for TCP transport highlights the practical importance of solutions like QNAME-based fragmentation~\cite{rawat2023post} or out-of-band key exchange mechanisms~\cite{beernink2022taking} for addressing packet size constraints.

Testing in production environments represents an important research direction. While this controlled laboratory evaluation provides valuable performance baselines, the results contrast with the large-scale RIPE ATLAS network testing conducted by Goertzen et al.~\cite{PQCexpRIPE}, highlighting the need for production deployment studies to validate these findings under real-world conditions. The controlled laboratory conditions used in this evaluation may not reflect the performance characteristics observed in actual DNS deployments, particularly in Kubernetes clusters where CoreDNS serves as the default DNS server. Production testing would provide insights into how post-quantum DNSSEC performs under actual traffic loads and network conditions that are more susceptible to issues affecting DNS resolution.

Finally, the integration of Stateful Hash-Based Signatures (STFL HBS) awaits implementation support in the Go programming language. These algorithms require state management since each private key can only be used once, and they use different signing functions that are not yet available in Go. Once language support becomes available, they could be added to CoreDNS following the same approach as in this project.

\section{Code Availability}

This implementation builds upon several established open-source projects and libraries. The complete source code and implementation artifacts are made available under open-source licenses to facilitate reproducibility and future research in post-quantum DNSSEC implementations.

\subsection{Implementation Repositories}

The main contributions of this work are available in the following repositories:

\begin{itemize}
\item Modified miekg/dns library with PQC algorithm support:
\url{https://github.com/qursa-uc3m/dns}~\cite{gentosuela2025dnspqc}
\item CoreDNS \texttt{dnssec\_pqc} plugin, version v0.1.1:
\url{https://github.com/qursa-uc3m/dnssec_pqc_plugin}~\cite{gentosuela2025plugin}
\end{itemize}

\subsection{Dependencies and Version Information}

The implementation was developed using the following software versions:

\begin{itemize}
\item CoreDNS: v1.12.2~\cite{coredns2025, coredns2025git}
\item Go: go1.23.11 linux/amd64
\item liboqs: 0.14.0-rc1~\cite{liboqs}
\item liboqs-go bindings~\cite{liboqsgo}
\item miekg/dns: v1.1.62~\cite{miekgdns2025} (base version for modifications)
\end{itemize}

\section*{Acknowledgments}
This work was supported by the Spanish Government under the following grants funded by MICIU/AEI/10.13039/501100011033: (i) ``QUantum-based ReSistant Architectures and Techniques (QURSA)" TED–2021–130369B–C32 and by the "European Union NextGenerationEU/PRTR"; and (ii) ``DIstributed Smart Communications with Verifiable EneRgy-optimal Yields (DISCOVERY)" PID2023-148716OB-C32. In addition, it was partially supported by the I-Shaper Strategic Project (C114/23), due to the collaboration agreement signed between the Instituto Nacional de Ciberseguridad (INCIBE) and the UC3M; this initiative is being carried out within the framework of the Recovery, Transformation and Resilience Plan funds, funded by the European Union (Next Generation). It was also supported by the Comunidad de Madrid under the grant ``RAMONES-CM" TEC-2024/COM-504.

\printbibliography

\end{document}